\documentclass[reprint,showpacs,preprintnumbers,amsmath,amssymb,prb]{revtex4-1}

\usepackage{graphicx}
\usepackage{color}

\begin{document}

\title{
High-temperature oxide thermoelectrics
}

\author{
Ichiro Terasaki
}

\email[Email me at:]{
terra@cc.nagoya-u.ac.jp
}
 
\affiliation{
Department of Physics, Nagoya University, 
Nagoya 464-8602, Japan
}

\begin{abstract}
We have evaluated the power factor of transition metal
oxides at high temperatures using the Heikes formula
and the Ioffe-Regel conductivity.
The evaluated power factor 
is found to be nearly independent of carrier concentration 
in a wide range of doping,
and well explains the experimental data for cobalt oxides.
This suggests that the same power factor
can be obtained with a thermopower larger than $2k_B/e$,
and also suggests a reasonably high value of
the dimensionless figure of merit $ZT$.
We propose an oxide thermoelectric power generator
by using materials having a thermopower larger than 300 $\mu$V/K.
\end{abstract}


\maketitle

\section{Introduction}
Thermoelectrics is a technology that can convert heat into electricity
and vice versa through the thermoelectric phenomena in semiconductors
and metals.\cite{mahan1998}
Owing to the pressing needs for reduction of the carbon-dioxide emission,
thermoelectric power generation has received a renewed interest from the
viewpoint of waste-heat recovery into electric power.
In particular, power generator at high temperature in air is 
highly desirable for waste heats from car engines
and garbage furnaces.
For such purposes, thermoelectric devices using oxide ceramics 
are quite promising, but the thermoelectric conversion
efficiency is not satisfactory at present.

After the discovery of the good thermoelectric properties
of the layered cobalt oxide Na$_x$CoO$_2$,\cite{terasaki1997}
oxide thermoelectrics using transition-metal oxides
is extensively investigated for the last 
decade.\cite{terasaki2003, maignan2002, maignan2002ce, %
koumoto2006}
Compared with other thermoelectric materials, transition-metal
oxides are quite unique in the sense that 
(i) the electron correlation is strong enough to make
the Heikes formula valid for the thermopower
above room temperature,\cite{chaikin1976,palstra1997,koshibae2000}
(ii) the resistivity is barely metallic in which the mean free
path is of the order of the unit cell length, and
(iii) the spin fluctuation significantly affects the 
physical properties.\cite{limelette2006,bobroff2007}
As a result, the thermoelectric parameters
cannot be predicted from conventional semiconductor physics.\cite{mahan1998}

Here we try to clarify how the power factor $S^2\sigma$ 
changes with carrier concentration and temperature
in transition-metal oxides,
where $S$ and $\sigma$ are the thermopower and the conductivity, 
respectively.
Although the thermopower of many transition-metal oxides
have been discussed with the Heikes formula,\cite{palstra1997,maignan2002ce,hebert2007}
there are only a few discussions on how 
the conductivity is described in the same 
conditions.\cite{kapitulnik1992,palsson1998,durczewski1998,koshibae2001} 
In particular, we wish to address the following two anomalies.
First, an optimum carrier concentration $n_0$
that maximizes the power factor
is of the order of 10$^{21}$ cm$^{-3}$,
and depends strongly on temperature.
Second, while the thermopower at $n_0$ often depends on materials,
the maximized power factor is approximately 2-3 $\mu$W/cmK$^2$,
and is nearly independent of material species.
These are seriously incompatible with the 
``collective wisdom'' of thermoelectrics, 
where $n_0$ is of the order
of 10$^{19}$ cm$^{-3}$, and
the optimized thermopower is approximately 2$k_B/e$.
To see these two anomalies we will take
La$_{1-x}$Sr$_x$CoO$_3$ as a typical example.
The doped LaCoO$_3$ has been extensively investigated 
as a possible candidate for thermoelectric 
oxide,\cite{androulakis2003,androulakis2004,robert2007,jirak2008}
and the thermoelectric parameters have been 
systematically measured in a wide range of carrier
concentration.\cite{masuda2003,kriener2004,iwasaki2008}
Iwasaki et al. \cite{iwasaki2008} have measured 
the thermoelectric parameters
using sixteen polycrystalline samples of La$_{1-x}$Sr$_x$CoO$_3$
from $x=0$ to 0.40 in a wide range of temperature from 77 to 1100 K.
Their comprehensive study offers a good example to test 
the idea discussed in the present paper.  
We discuss a possible value of the figure of merit 
for such materials at high temperatures.

\section{Theoretical background}
Let us briefly summarize the conventional thermoelectrics
by following Mahan's review article.\cite{mahan1998}
According to the Boltzmann equation, the thermopower $S$
can be written with the Fermi-Dirac distribution $f_0$
and the chemical potential $\mu$ by
\begin{equation}
S  = \frac{1}{eT}\frac{\int
{\left({-\frac{\partial f_0 }{\partial \varepsilon }}
\right)_{\varepsilon =\varepsilon_{\bf k}} 
v_{\bf k}^2\tau_{\bf k} (\varepsilon_{\bf k}-\mu)d^3k}}
{\int {\left( {-\frac{\partial f_0 }{\partial \varepsilon }} 
\right)_{\varepsilon =\varepsilon_{\bf k}} 
v_{\bf k}^2\tau_{\bf k} d^3k}},
\label{calc_S}
\end{equation}
where $\varepsilon_{\bf k}$, $v_{\bf k}$ and $\tau_{\bf k}$ are 
the energy,  velocity
and scattering time for an electron with a wavevector of {\bf k},
respectively.
We can also express the carrier density $n$ as
\begin{equation}
n = 2\int \frac{d^3k}{(2\pi)^3} f_0(\varepsilon_{\bf k}).
\label{calc_n}
\end{equation}
Let us evaluate Eqs. (\ref{calc_S}) and (\ref{calc_n}) at high 
temperatures. 
The Fermi-Dirac distribution function $f_0$
can be approximated as the Maxwell-Boltzmann distribution as
\begin{equation}
f_0=\frac{1}{\exp(x-\beta\mu)+1} \sim \exp(\beta\mu -x),
\end{equation}
where $\beta =1/k_BT$ and $x=\varepsilon_{\bf k}/k_BT$.
Now we can evaluate the integrals in Eqs. (\ref{calc_S}) and (\ref{calc_n}),
by assuming a single band with a parabolic dispersion.
Then we get $n$ as
\begin{equation}
n = \frac{1}{\pi^2}\left(\frac{2mk_BT}{\hbar^2}\right)^{3/2}
\frac{\sqrt{\pi}}{2}\exp(\beta\mu)
\end{equation}
For many practical cases, this equation determines the chemical potential $\mu$
as a function of $n$ and $T$ as
\begin{equation}
\mu = k_BT\ln\left(\frac{n}{\xi}\right),
 \label{mu}
\end{equation}
where
\begin{equation}
\xi\equiv \frac{1}{\pi^2}\frac{\sqrt{\pi}}{2}
\left(\frac{2mk_BT}{\hbar^2}\right)^{3/2}.
\end{equation}

In the same way, we evaluate the thermopower as
\begin{eqnarray}
S &=& -\frac{k_B}{e}\frac{\int d^3k\ v_{\bf k}^2\tau_{\bf k} 
(\beta\mu-x)\exp(\beta\mu-x)}
{\int d^3k\ v_{\bf k}^2\tau_{\bf k}\exp(\beta\mu-x)}\nonumber\\
&=& -\frac{k_B}{e}(\beta\mu -\langle x\rangle),\nonumber\\
&=& -\frac{k_B}{e}\left(\ln\left(\frac{n}{\xi}\right) -\langle x\rangle\right)
\label{hightempS}
\end{eqnarray}
where $\langle x\rangle$ is the average of $x$ with a weight function of
$v_k^2\tau_{\bf k}\exp(\beta\mu-x)$. 
Note that we used Eq. (\ref{mu}) to eliminate $\mu$.
Now we can get the optimum condition of the power factor
by differentiating it with $n$.
By neglecting the carrier-concentration dependence
of the mobility, we get
\begin{equation}
\frac{d(S^2\sigma)}{dn}
\propto \frac{d\ }{dn}\Bigl[n(\ln(n/\xi)-\langle x\rangle)^2\Bigr]
=0.
\label{maximizePF}
\end{equation}
Then the optimum carrier concentration $n_0$ should
satisfy
\begin{equation}
\ln\left(\frac{n_0}{\xi}\right)=\langle x\rangle-2. 
\label{n0}
\end{equation}
By eliminating $\langle x\rangle$ from Eq. (\ref{hightempS}),
we get the optimum thermopower $S_0$ as
\begin{equation}
S_0 = \frac{k_B}{e}
\left(-\ln\left(\frac{n_0}{\xi}\right)+\langle x\rangle\right)
= \frac{2k_B}{e}
\label{s0}
\end{equation}
This indicates that \textit{the optimum thermopower
is universal and independent of materials parameters}.
Using realistic parameters for $x$, we can evaluate $n_0$
to be of the order of $10^{19}$ cm$^{-3}$ at 300 K.
In fact, the conventional thermoelectric materials 
are optimized around this carrier concentration,
and show a thermopower of $\pm$ 200 $\mu$V/K
that is close to $\pm 2k_B/e=\pm$170 $\mu$V/K.
Very recently, Pichanusakorn and Bandaru \cite{pichanusakorn2009}
have performed a similar but more extensive calculation,
and have found that the optimized thermopower is always
in a limited range of 130--187 $\mu$V/K, which is essentially independent
of material parameters.

\begin{figure}
\begin{center}
 \includegraphics[width=8cm,clip]{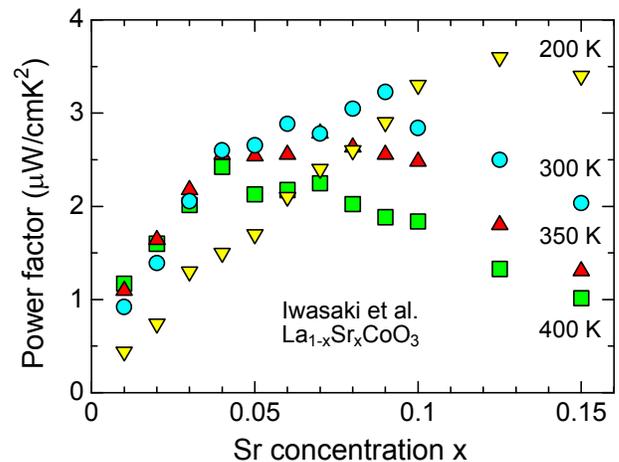}
\end{center}
\caption{(Color online)
The power factor of polycrystalline samples of 
La$_{1-x}$Sr$_x$CoO$_3$ measured by Iwasaki et al.\cite{iwasaki2008}
plotted as a function of Sr concentration $x$.
}
\label{fig1}
\end{figure}

\section{Power factor for disordered systems}
According to Eq. (\ref{n0}),
$n_0$ is proportional to $T^{3/2} \exp(\langle x\rangle -2)$,
meaning that the optimum carrier concentration is 
higher for higher temperatures. 
This is  not true in real transition-metal oxides.
Figure 1 shows the power factor
of polycrystalline samples of La$_{1-x}$Sr$_x$CoO$_3$
measured by Iwasaki et al. \cite{iwasaki2008}
as a function of Sr concentration $x$
for several temperatures.
The optimum Sr concentration $x_0$
depends clearly on temperature.
This is reasonable, because Eq. (\ref{maximizePF}) 
is based on the assumption that
the mobility is independent of carrier concentration.
In real materials, however,  the mobility
is highly dependent on carrier concentration
especially near metal-insulator transitions.

Let us have a closer look at the temperature dependence
of the optimum Sr concentration $x_0$.
While $x_0$ is around 0.12 (corresponding to 
$n=$2$\times$10$^{21}$ cm$^{-3}$) at 200 K, 
$x_0$ shifts to lower $x$ with increasing temperature.
The power factor at 350 K shows a plateau from $x=0.04$ to 0.1,
which is understood from two broad maxima located 
near $x=$0.04 and 0.1.
In this context, $x_0$ gradually decreases with temperature 
from $x_0=$0.12 at 200 K, through 0.1 at 300 K, to 0.04 at 400 K.

Let us discuss an origin for the temperature
dependence of $x_0$.
In La$_{1-x}$Sr$_x$CoO$_3$, the thermopower
from 200 to 400 K are weakly dependent on temperature.
As is widely known, the temperature-independent thermopower 
has been analyzed using the Heikes formula.\cite{chaikin1976}
In this formula, entropy per site is associated with the thermopower.
In a simplest case, the thermopower of the Heikes formula $S_{\rm H}$
can be given by
\begin{equation}
 S_{\rm H} = \frac{k_B}{e}\ln\frac{2x}{1-x},
\label{heikes}
\end{equation}
where $x$ is the carrier concentration per unit cell.
This formula is valid, when the thermal energy $k_BT$
is much larger than the transfer energy $t$
but much smaller than the on-site Coulomb repulsion $U$
($t \ll k_BT \ll U$).
Although the spin and orbital degrees of freedom
exist in the Co ions, we neglect in the present paper
to avoid the controversy of the spin
states of the doped LaCoO$_3$.\cite{kriener2004,asai1994,wu2003}

In the same condition, we should employ
the conductivity for nonmetallic conduction.
One extreme case is known as the Ioffe-Regel limit,
where the electron mean free path is close to
the lattice parameters.
In this picture, the Ioffe-Regel conductivity 
$\sigma_{\rm IR}$ is given by
\begin{equation}
 \sigma_{\rm IR} = 0.33 x^{2/3}\frac{e^2}{\hbar a},
\label{IRcond}
\end{equation}
where $a$ is the lattice parameter.\cite{gurvitch1981}
Thus the power factor is written as
\begin{equation}
 S_{\rm H}^2 \sigma_{\rm IR}  \propto x^{2/3}\left[\ln\frac{2x}{1-x}\right]^2.
\label{PF1}
\end{equation}

Another extreme case is the electrical
conduction near the Anderson transition.
In this case, the conductivity $\sigma_{\rm A}$
goes toward zero critically as 
$\sigma_{\rm A} \propto (n-n_c)^{p}$,
when $n$ approaches the critical concentration $n_c$.
When the critical regime is sufficiently large,
we get $\sigma_{\rm A} \propto n^p$ for $n \gg n_c$.
In this case, the power factor is written as
\begin{equation}
 S_{\rm H}^2 \sigma_{\rm A} \propto x^{p}\left[\ln\frac{2x}{1-x}\right]^2.
\label{PF2}
\end{equation}

\begin{figure}
\begin{center}
 \includegraphics[width=8cm,clip]{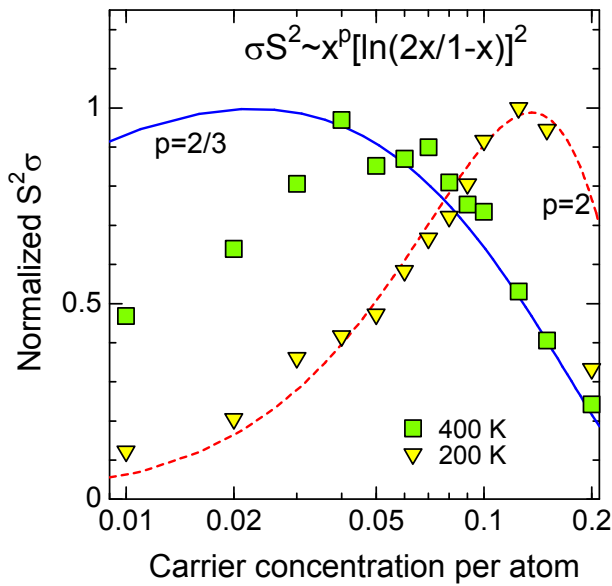}
\end{center}
\caption{(Color online)
The power factor calculated from
Eq. (\ref{PF1}) (solid curve) and Eq. (\ref{PF2}) (dotted curve).
plotted as a function of carrier concentration per unit cell.
The power factors at 200 and 400 K in Fig.1 are
also plotted.
}
\label{fig2}
\end{figure}

Figure 2 shows the normalized power factors calculated from
Eqs. (\ref{PF1}) and (\ref{PF2}), together with 
the experimental data shown in Fig.1,
where the properly normalized values of 
$x^{p}\left[\ln2x/(1-x)\right]^2$ 
is plotted for $p=2/3$ [Eq. (\ref{PF1})]
and for $p=2$ [Eq. (\ref{PF2})] by the solid and dotted curves,
respectively.
As is clearly seen by the dotted curve in Fig. 2,
the exponent of $p=2$ well explains the power factor measured at 200 K.
This suggests that the mobility is proportional to $x$,
which is consistent with 
the mobility at 300 K evaluated by Iwasaki et al.\cite{iwasaki2008}
This doping dependent mobility is the reason why 
the optimum carrier concentration
is much larger than the value given by Eq. (\ref{n0}).
At higher temperature, in contrast, all the samples tend to 
show $\sigma_{\rm IR}$ called resistivity saturation,\cite{gurvitch1981} and 
the solid curve explains the measured data at 400 K for $x>0.03$. 
It is reasonable that the measured data is below the solid curve 
for $x\le 0.03$, because at such low carrier concentration,
strong localization takes place to make the conductivity lower
than $\sigma_{\rm IR}$.

Looking at Eqs. (\ref{PF1}) and (\ref{PF2}), one can understand why 
the thermoelectric performance of the 
thermoelectric oxides is poor at room temperature, but
is comparable to that of other materials above 1000 K.
The low-temperature mobility is poorer for lower carrier concentration,
and the conductivity $\sigma_{\rm A}$ is much lower than that
of the conventional thermoelectric materials.
On the other hand, when Eq. (\ref{PF1}) is valid,
the resistivity and thermopower become independent of temperature.
For $x=0.02$ and $a=5$\AA, for example, 
we have $\sigma_{\rm IR}=$100 S/cm and $S_{\rm H}=$300 $\mu$V/K.
Then the power factor equals 9 $\mu$W/cmK$^2$.
Note that the value of 9 $\mu$W/cmK$^2$ is
the upper limit for materials in which the Heikes formula is valid.
It is thus reasonable to regard
this value  as close to the observed maximum power factor of 
2-3 $\mu$W/cmK$^2$ in Fig. 1.
Assuming a low thermal conductivity $\kappa$ of 10 mW/cmK, 
we estimate the figure of merit $Z=S_{\rm H}^2\sigma_{\rm IR}/\kappa$ to be
9$\times$10$^{-4}$ K$^{-1}$, which gives $ZT=0.9$ at 1000 K.
The thermoelectric parameters of the layered cobalt oxides roughly 
meet the above conditions.\cite{ohtaki2000,fujita2001,funahashi2002,shikano2003,ito2003,nong2011}
A prime example is Ca$_3$Co$_4$O$_9$;
At 1000 K, a single crystal shows a power factor of 26 $\mu$W/cmK$^2$
with a resistivity of 2.4 m$\Omega$cm and a thermopower
of 250 $\mu$V/K,\cite{shikano2003} and
a ceramic sample shows a power factor of 6 $\mu$W/cmK$^2$
with a resistivity of 9 m$\Omega$cm and a thermopower 
of 230 $\mu$V/K.\cite{nong2011}
We should further emphasize that the above estimation is
done \textit{in the limit of low mobility}. 
This is also incompatible
against the collective wisdom of thermoelectrics, 
where high-mobility semiconductors are required for high $ZT$ values.
Equation (13) warrants that highly disordered materials can
be fairly good thermoelectric materials at high temperature.

Another feature of Eq. (\ref{PF1}) is that 
the solid curve in Fig. 2 indicates that the power factor has a broad
maximum around 0.01-0.06.
This is because $\sigma_{\rm IR}$ is weakly dependent 
on $x$ so that the increase in $S_{\rm H}$ compensates 
the decrease in $\sigma_{\rm IR}$ well.
Equation (\ref{heikes}) gives 340 and 170 $\mu$V/K for $x=0.01$
and 0.06, respectively, 
and materials with a thermopower larger than 
300 $\mu$V/K should be designed to maximize the power factor.
In fact, the thermopower for $x=0.03$ is 300 $\mu$ V/K at 400 K.
This is again seriously incompatible with the prediction of
Eq. (\ref{s0}) that 
the optimized thermopower is always $2k_B/e=$170 $\mu$V/K.

Kobayashi et al.\cite{kobayashi2007ict} have reported
that the power factor of the misfit layered cobalt oxides
[Bi$_2$A$_2$O$_4$][CoO$_2$]$_{b_1/b_2}$
is nearly independent of materials at 300 K.
By changing the block layer, they controlled 
the carrier concentration, and concomitantly 
changed the resistivity and thermopower from sample to sample.
An important finding is that the power factor at 300 K is close
to 2 $\mu$W/cmK$^2$ for all the samples measured. 
We think this doping-independent power factor 
is essentially understood with Eq. (\ref{PF1}).

The present work suggests that the thermopower of transition-metal oxides 
can be increased up to 340 $\mu$V/K with remaining the power factor unchanged,
when the transport parameters are explained with the Heikes formula.
By using such oxides, one can reduce the number of thermocouple 
without deteriorating the performance of the device.
Such devices are easier to fabricate, easier to lower contact resistance,
and easier to take impedance matching because of its high impedance.
We hope that we can increase the yielding ratio 
and the reliability as well by using such devices.
Recently, Bonetti et al. \cite{bonetti2011} have reported 
that a huge thermopower of 7 mV/K around 30-45$^{\circ}$C 
for some electrolytes, and argued the same idea.

\section{Summary}
We have discussed the high-temperature power factor 
of disordered materials using the 
Heikes formula for the thermopower
and the Ioffe-Regel limit for the conductivity.
We have found that this power factor can be as large as 
9 $\mu$W/cmK$^2$ (a resistivity of 10 m$\Omega$cm
and a thermopower of 300 $\mu$V/K)
at any temperatures, which could give
a reasonable value of $ZT=0.9$ at 1000 K with a
realistic value of the thermal conductivity of 10 mW/cmK.
We have further found that this power factor takes a 
broad maximum for a wide range of carrier concentration
from 0.01  to 0.06 per unit cell.
Since this carrier concentration range corresponds to 
a thermopower from 170 to 340 $\mu$V/K,
a thermopower larger than 300 $\mu$V/K is applicable for
a thermoelectric power generator using transition-metal oxides.

\section*{Acknowledgments}
The author wish to thank Y. Horiuchi, D. Sawaki, A. Inagoya
for collaboration on experimental searches for 
large-thermopower materials,
and also appreiate W. Kobayashi and S. H\'ebert for 
stimulating input about the doping-independent power factor.
This work was partially supported 
by Tokuyama Science Foundation, and
by Strategic Japanese-Finland Cooperative Program 
on ``Functional Materials'', JST, Japan.


%

\end{document}